\documentclass{PoS}
\usepackage{amsbsy, amsmath,mathrsfs}
\usepackage{hhline}
\usepackage{subfigure}

\newcommand{\bea}{\begin{eqnarray}}
\newcommand{\eea}{\end{eqnarray}}
\newcommand{\bean}{\begin{eqnarray*}}
\newcommand{\eean}{\end{eqnarray*}}
\def\l{\left}
\def\r{\right}
\newcommand{\gsw}{{\mathscr G_s}(w)}
\newcommand{\gsone}{{\mathscr G_s}(1)}
\newcommand{\gw}{{\mathscr G}(w)}

\newcommand{\pDs}{\ensuremath{p_{_{\!D_s^{}}}}}%
\newcommand{\pBs}{\ensuremath{p_{_{\!B_s}}}}%
\newcommand{\gev}{{\rm GeV}}

\title{Lattice QCD study of ${B_s\to D_s \ell \bar\nu_\ell}$ decay near zero recoil}

\ShortTitle{$B_s\to D_s$ near zero recoil}

\author{\speaker{Mariam Atoui}\thanks{CNRS Liban Scholarship}, Vincent Mor\'enas \\
        Laboratoire de Physique Corpusculaire LPC, Universit\'e Blaise Pascal\\
        E-mail: \email{matoui@in2p3.fr}, $\,$ \email{morenas@in2p3.fr} }

\author{{Damir Be\v{c}irevi\'c}, Francesco Sanfilippo \\
        Laboratoire de Physique Th\'eorique, Universit\'e Paris Sud\\
        E-mail: \email{damir.becirevic@th.u-psud.fr},  \email{fr.sanfilippo@gmail.com}}

\abstract{We study the hadronic matrix elements describing the $B_s \to D_s\ell \bar\nu_\ell$ decay in and beyond the Standard Model. By using the twisted mass
QCD on the lattice with $N_f=2$ dynamical flavors we compute the normalization $\mathscr G_s(1)$
of the form factor dominating $B_s\to D_s \ell \bar\nu_\ell$ in the SM.
We also make the first lattice determination of $F_0(q^2)/F_+(q^2)$ and $F_T(q^2)/F_+(q^2)$ near zero recoil (near $q^2_{\rm max}$).
We briefly discuss the non-strange case $B \to D\ell \bar\nu_\ell$ as well. }

\FullConference{31st International Symposium on Lattice Field Theory - LATTICE 2013\\
		July 29 - August 3, 2013\\
		Mainz, Germany}

\begin{document}
\section{Introduction}
\small
\noindent A precise knowledge of semileptonic decays of $B$-mesons brings several advantages to flavor physics.
For example, the decay channel $B\to D \ell \bar\nu_\ell$ 
%and $B \to D^* \ell \bar\nu_\ell$
allows an independent estimate of the CKM matrix element ($V_{cb}$)
which is extracted by comparing the theoretical determination of form factors
with experimental measurements of the partial or total decay widths. 
In the limit of vanishing lepton mass, the differential decay rate of $B \to D \ell \bar\nu_\ell$ reads~\cite{manohar} 
%and $B \to D^* \ell \bar\nu_\ell$ write~\cite{manohar} 
\bea
\label{exclusiveff}
\dfrac{d\Gamma}{dw} ({B}\to D \, \ell \bar\nu_\ell) \, &= \, \dfrac{G_F^2}{48\pi^3} \, (m_B+m_D)^2 \, m_D^3 (w^2-1)^{3/2} \, |V_{cb}|^2 \, |\mathscr{G}(w)|^2 \,,
\eea
where $\mathscr{G}(w)$ is the relevant form factor and $w$ is the product of the velocities of the hadrons ($w=v_B\cdot v_D$) in the HQET framework.\\
\noindent The main theoretical problem is the knowledge of $\gw$.
At the zero recoil point, $w=1$, heavy quark symmetries play a useful role in setting $\mathscr{G}(1)=1$ in the limit of $m_{b,c}\to \infty$~\cite{isgurwise}.
Together with the short distance QCD corrections, the effect of the finiteness of $m_b$ and $m_c$ masses leads to the fact that $\mathscr{G}(1)\neq1$, which is a non-perturbative effect that needs to be computed by means of lattice QCD.\\
%However, this point is not directly accessible from experiments so that an additional theoretical input (the zero recoil point $\gone$ or the Isgur-Wise point) is needed to extract $|V_{cb}|$.\\

\noindent Here we propose to study the $B_s\to D_s$ transitions in and beyond the Standard Model (SM).
The decay mode $\bar B_s^0\to D_s^+\ell \bar\nu_\ell$ could be studied at LHCb and especially at Super Belle.
One more advantage in studying $B_s$ decay is that the soft photon problem is less important than in the non-strange case in which the charged and neutral $B$-semileptonic decay modes are averaged~\cite{Becirevic:2009fy}.
From the lattice point of view, the non-strange heavy-light mesons are more difficult because a chiral extrapolation in the valence light quark
is required, which is a source of systematic uncertainties. Working with the strange case is simpler because the light spectator is fixed to its known mass $(m_s)$ and no extrapolation in the light quark mass is needed when computing the form factors on the lattice.\\

\noindent In the following, we first introduce the relevant form factors contributing to the semileptonic decays of $B_s$ into $D_s$ mesons 
in the SM. We then proceed to describe the simulation details as well as the strategy for our computation. We present our results at 
the zero recoil point for $\mathscr G_s(1)$ and then we look for New Physics (NP) beyond the SM in $B_s\to D_s \ell \bar\nu_\ell$ decays
by introducing the scalar and the tensor form factors. More details can be found in Ref.~\cite{Atoui:2013zza}.

\section{$\boldsymbol{B_s\to D_s}$ form factors}
\noindent The hadronic matrix element governing the $B_s \to D_s$ decay is parametrized in the SM as
% by the vector and the scalar form factors, $F_+(q^2)$ and $F_0(q^2)$, as
\bea\label{eq:def1}
\begin{aligned}
\langle D_s(\pDs) | V_\mu | \bar B_s(\pBs) \rangle  =
{{\ F_+(q^2)}}\,(\pBs+\pDs)_{\mu} 
+ q_\mu \; \left[ {{F_0(q^2) - F_+(q^2) }} \right ] \left( \dfrac{m_{B_s}^2-m_{D_s}^2}{q^2} \right)\,,
\end{aligned}
\eea
where the vector ($F_+(q^2)$) and the scalar ($F_0(q^2)$) form factors 
are functions of $q^2=(\pBs-\pDs)^2$, that can vary within the range $q^2 \in [ m_\ell^2, q_{\rm max}^2]$, where $q_{\rm max}^2=(m_{B_s}-m_{D_s})^2$.\\
We choose to work in the $B_s$ rest frame and give a momentum to the $D_s$ meson. We will take this momentum to be symmetric in its spatial components 
%\bean
$\pBs = \left(m_{B_s} \, , \, \vec 0 \right) \; {\rm and} \;\pDs \,= \left( E_{D_s},p,p,p \right)$.
%\eean
We also use the twisted boundary conditions (BCs)~\cite{nazario} for the quark field.
%in order to increase the kinematical region accessible for the investigation of form factors.
This allows to shift the quantized values of $p$ by a continuous amount
 %\bean
$ p= \dfrac {\theta \pi}{  L} \; {\text{ so that}}  \; |\vec q|= \sqrt{3} \, \dfrac{\theta\pi} { L}\,.$
% \eean
The $\theta$'s we choose correspond to small momenta, thus we are discussing the decay matrix element near zero-recoil (near $q^2_{\rm max}$). More specifically,
%The resulting $D_s$ obey the free-boson lattice dispersion relation. 
our $\theta$'s correspond to the following recoils $w$ for $B_s \to D_s \ell \bar\nu_\ell$:
\bea
\label{nonzero}
w \in \, \{1, \,  1.004,  \,1.016, \,1.036, \,1.062 \}\,.
\eea
Note that $w_{\rm max}$ for this decay mode, in the case of massless leptons, is 1.546.\\
One can separate the contribution proportional to the scalar form factor $F_0(q^2)$
 \bea
% \l\{
% \begin{aligned}
P_\mu^0 \langle D(p_{D_s})\vert V_\mu\vert B(p_{B_s})\rangle &= \dfrac{m_{B_s}+m_{D_s}}{ m_{B_s}-m_{D_s}}\; F_0(q^2)\; \;
%\,, \\[2mm]
{\rm with} \;P_\mu^0 &= {q_\mu\over q^2_{\rm max}}, \; \; {\rm and} \; q_\mu=(m_{B_s}-E_{D_s}, -\vec p_{D_s})\,,
%\end{aligned}
%\r.
\eea
and that of the vector form factor $F_+(q^2)$ as
\bea
\label{projectorplus}
%\l\{
%\begin{aligned}
 P_\mu^+ \langle D(p_{D_s})\vert V_\mu\vert B(p_{B_s})\rangle &= {\vec q\ }^2 \dfrac{2 m_{B_s}}{ m_{B_s} - E_{D_s}} \; F_+(q^2)\;
 %,\\
 {\rm with} \; P_\mu^+ &= \left({{\vec q\ }^2\over m_{B_s}-E_{D_s}},\vec q \right)\,.
% \end{aligned}
 %\r.
 \eea
%%%%%%%%%%%
\section{Lattice strategy and setup}
\noindent In this analysis, we use the gauge field ensembles produced by the European Twisted Mass 
Collaboration~\cite{Boucaud:2007uk,Boucaud:2008xu} at four values of the lattice spacing corresponding to four values of the inverse bare gauge coupling $\beta$.
Dynamical quark simulations have been performed using the tree-level improved Symanzik gauge action~\cite{Weisz:1982zw}
and the Wilson twisted mass quark action~\cite{Frezzotti:2000nk} tuned to the maximal twist~\cite{Frezzotti:2003ni}.
Bare quark mass parameters, corresponding to a degenerate bare mass value of the $u/d$ quark, 
are chosen to have the light pseudoscalar mesons (PS) in the range 
$280 \leq m_{\rm PS} \leq 500$~MeV. The details of the simulation parameters are collected in Ref.~\cite{Atoui:2013zza}.\\
We computed all the quark propagators by using stochastic sources, and then applied the so-called one-end trick to compute the needed correlation functions~\cite{Boucaud:2008xu}.
%Two- and three-point correlation functions were then computed by employing smearing techniques.\\
We work with ten heavy quark masses $m_h$ starting from the charm quark mass, $m_h^{(0)}=m_c$, and then successively 
increase the heavy quark mass by a factor of $\lambda=1.176$, $m_h^{(i)}=\lambda^i m_c$, so that after $9$ steps one arrives at $m_h^{(9)}=m_b=\lambda^{9}m_c$.\\
In order to extract the form factors $F_0(q^2)$ and $F_+(q^2)$ from lattice data, we compute the two- and three-point correlation functions, $\mathscr C^{(2)}_H(t)$ and $\mathscr C_\mu^{(3)}(\vec q,t)$ and, from which we then
extract the desired hadronic matrix elements as
\[
\mathscr C^{(3)}_\mu(\vec q;t) \xrightarrow{0 \; \ll \; t \ll \; t_S}  \dfrac{\mathscr Z_{B_s} \mathscr Z_{D_s}}{4m_{B_s} E_{D_s}} \exp\l( -m_{B_s} t\r) \times
\exp \l[-E_{D_s} (t_S-t) \r] \langle D_s(\vec k) | V_\mu | B_s(\vec 0) \rangle\,,
\]
where we fix $t_S=T/2$ ($T$ is the size of the temporal extension of the lattice). $m_{H}$, $E_{H}$ and $\mathscr Z_{H}$ are extracted from the large time behavior of the two-point correlation functions,
\[
\mathscr C^{(2)}_H(t) \xrightarrow{t \gg 0} \l| \mathscr Z_H \r|^2 \dfrac{\cosh \l[ m_H (T/2-t )\r]}{m_H} \exp \l[-m_H T/2 \r]\,,
\]
where $H$ stands for either $D_s$ or $B_s$ meson. In the computation of
correlation functions we applied the smearing procedure on the source operators (see~\cite{Atoui:2013zza} for details).
%already
%discussed in~\cite{charm2}. 
%The smeared
%field is of the form
%\[
%S^{(N)}  = (1+6\kappa_s)^{-N}
 %(1 + \kappa_{\rm{s}} a^2 \nabla^2_{\rm{APE}})^{N} S^{(0)}
%\]
%where $S^{(0)}$ is a standard local source and $\nabla_{\rm{APE}}$ 
%is the lattice covariant derivative with APE smeared gauge links characterized by the parameters  $\alpha_{\rm{APE}}=0.5$ and $N_{\rm APE}=20$. We have taken $\kappa_s=4$ and $N=30$.
%%%--------------------------------------
\section{Extraction of $\gsone$}
\begin{figure}[h!]
\subfigure[]
{
{\includegraphics[scale=0.5,angle=-0]{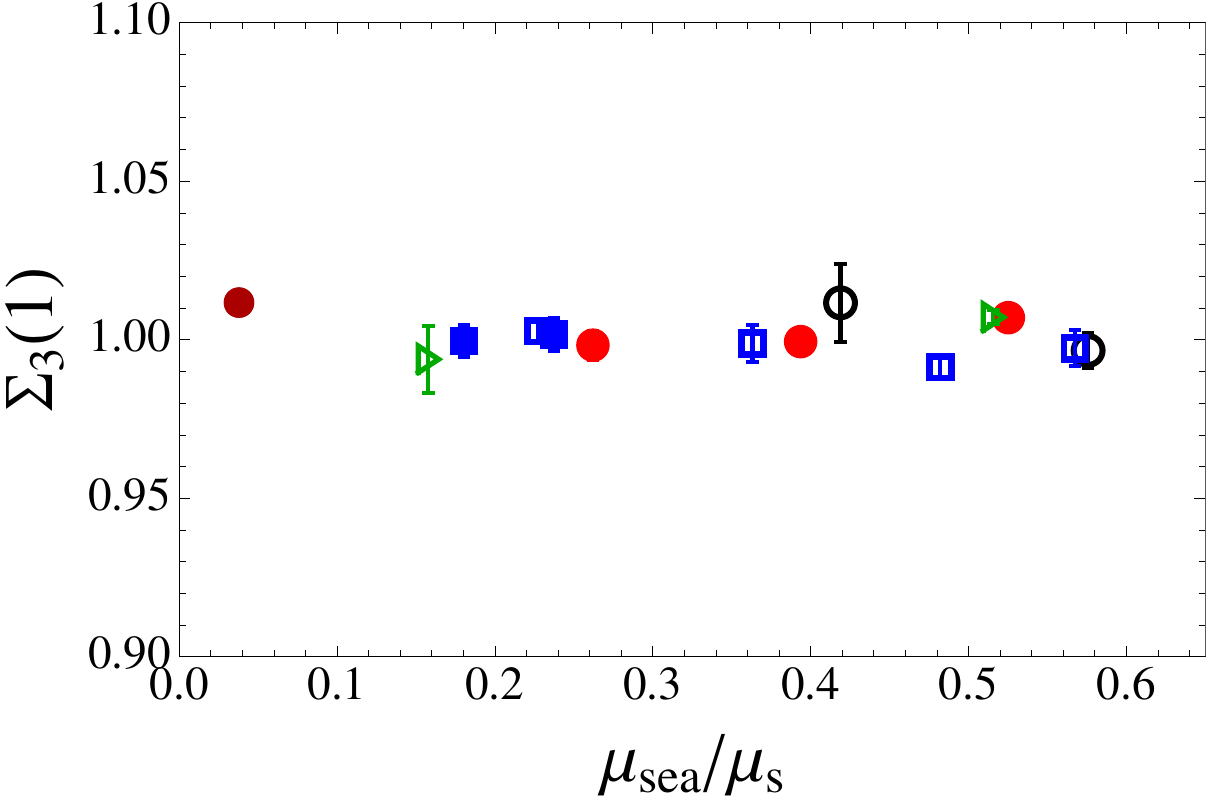}}
\label{fig:Sigma}
}
\hskip 1.5cm
\subfigure[]
{
{\includegraphics[scale=0.41,angle=-0]{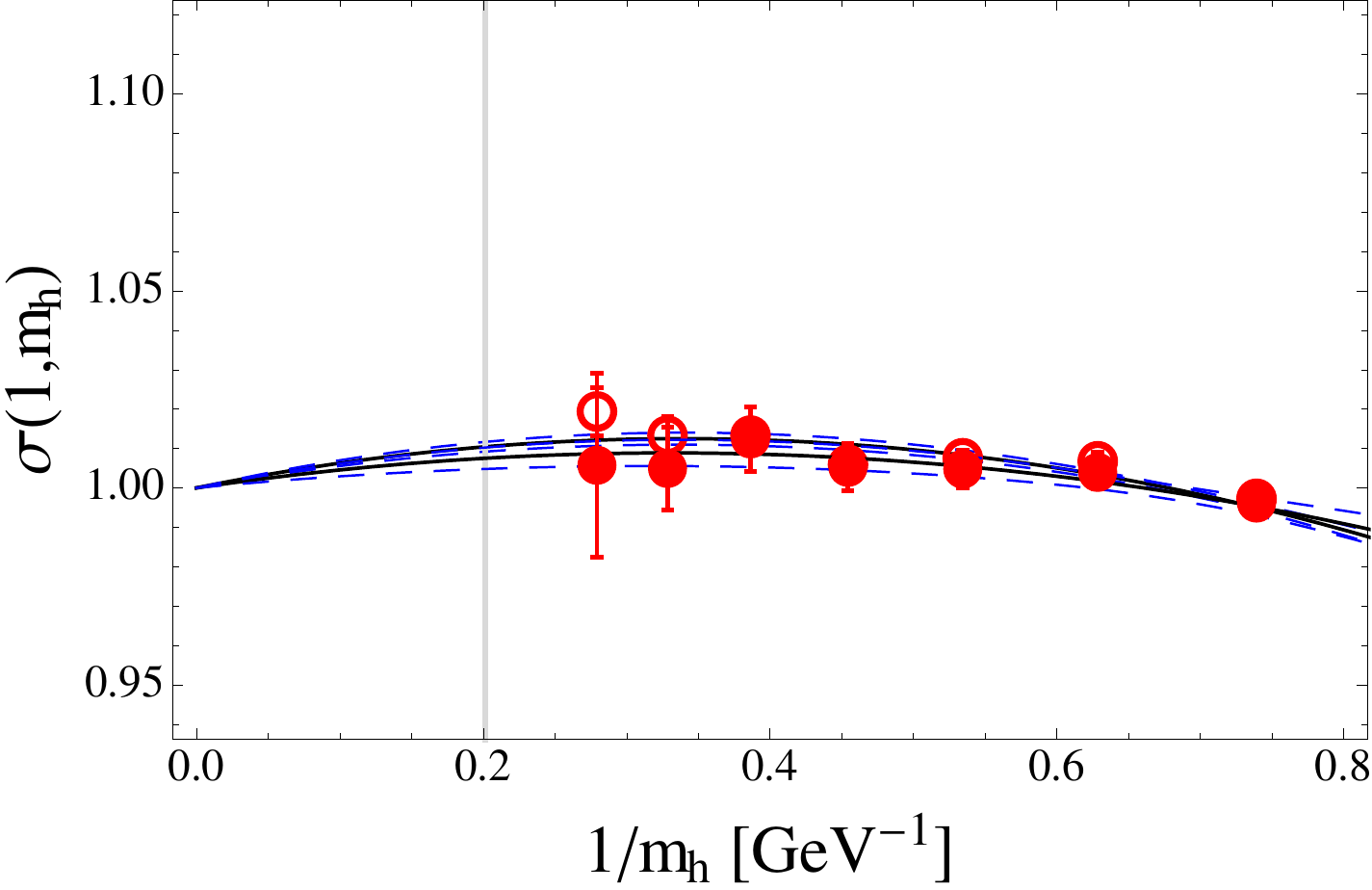}}
\label{fig:sigma}
}
\begin{center}
\vskip -0.5cm
\caption{\small \sl (a) Values of $\Sigma_3$
% obtained on all of the lattices used in this work, 
as a function of $\mu_{\rm sea}/\mu_s$.  {\Large{$\boldsymbol \circ$}} for $\beta=3.80$,  { \color{blue}$\boldsymbol\square$} for $\beta=3.90$ ($24^3$), ${\color{blue} \large \blacksquare}$  for $\beta=3.90$ ($32^3$),
 {\color{red} \LARGE \textbullet} for $\beta=4.05$,  and  ${\color{green} \boldsymbol{ \rhd}}$ for $\beta=4.20$. 
 The result of continuum extrapolation is indicated at $\mu_{\rm sea}/\mu_s = m_{ud}/m_s= 0.037(1)$.% ($m_{ud}$ is the average of the $u$ and $d$ quark masses).
 (b) Heavy quark mass dependence of the ratio $\sigma$ extrapolated to the physical value of the heavy quark mass. The vertical line 
represents the value of the inverse physical $b$ quark mass. 
Filled symbols correspond to $\sigma(1,m_h)$ extrapolated to the continuum limit with all parameters free, whereas the empty symbols refer to the results obtained by imposing $\beta_k^s=0$.}
\end{center}
\end{figure}

%%%--------------------------------------
\noindent Often used parametrization of the matrix element~\eqref{eq:def1} is the one motivated by the heavy quark effective theory (HQET) and reads
\bea\label{eq:def2}
\begin{aligned}
{1\over \sqrt{m_{B_s}m_{D_s}}} \langle D_s(v_{D_s})\vert V_\mu\vert B_s(v_{B_s})\rangle 
= \l( v_{B_s} + v_{D_s} \r)_\mu h_+(w) + \l( v _{B_s}- v_{D_s} \r)_\mu h_-(w)\,.
\end{aligned}
\eea
The desired $\gsw$ is then written as
\bean
\gsw \; = h_+(w) \left[ 1 - \;\l( {m_{B_s}-m_{D_s}\over m_{B_s}+m_{D_s}}\r)^2\, H(w) \right]\,, \quad \text{where} \quad H(w) =  \l( {m_{B_s}+m_{D_s}\over m_{B_s}-m_{D_s}}\r) \; {h_-(w)\over h_+(w)}\,.
\eean
To determine $\mathscr G_s$ at the zero recoil point, $w=1$, one needs to combine $h_+(1)$ with $H(1)$. The form factor $h_+(1)$
can be obtained from $F_0(q^2_{\rm max})$ as,
\[
h_+(1)\;= \dfrac{m_{B_s}+m_{D_s}}{\sqrt{4m_{B_s}m_{D_s}}}\,F_0(q^2_{\rm {max}})\, ,
\]
whereas $H(1)$ is not directly accessible from the lattice data. $H(w)$ is computed at
different values of $w$ (very close to $w=1$) and then extrapolated to $H(1)$.\\
In that way we computed $\gsone$ for each value of the heavy quark mass $m_h$. Note that the charm and strange quark masses are kept fixed.
We then form the ratios
\[
\Sigma_k(1,m_h^{(k)},a^2)\;=\dfrac{\mathscr{G}_s(1,m_h^{(k+1)},a^2)}{\mathscr{G}_s(1,m_h^{(k)},a^2)}\; ,
\]
where we indicate the dependence on the lattice spacing.\\
In Fig.~\ref{fig:Sigma}, we see that our lattice data exhibit very little or no dependence on the light sea quark mass, or on the lattice spacing. %Note also that for larger heavy quark masses the errors on $\sigma_k$ are larger, and therefore the corresponding continuum value $\sigma_k$ will have larger error as well.
%************************
%***********************
%**********************
%*********************
Each of the ratios is then extrapolated to the continuum limit
and to the physical sea quark mass by fitting the data to a linear function of $m_l^{\rm {sea}}$ and $a^2$,
\bea
\label{cont.extrap.}
\Sigma_k(1,m_h^{(k)},a^2) = \alpha^s_k +  \beta^s_k \; \dfrac{m_l ^{\rm sea}}{m_s}+ \gamma^s_k \; {a^2\over a_{3.9}^2}\;.
\eea
Since our data do not exhibit a dependence on the sea quark mass we also extrapolated $\Sigma_k(1)$ by imposing $\beta^s_k=0$
in Eq.~\eqref{cont.extrap.}. For higher masses, the results of the continuum extrapolation
have larger error bars in the case of a free $\beta^s_k$, as it can be seen in Fig.~\ref{fig:sigma}.\\
We identify $\sigma_k(1,m_h)$ with the continuum limit of $\Sigma_k(1,m_h^{(k)},a^2)$. 
Since $\displaystyle \lim_{m_h\to \infty}\gsone \to {\rm constant}$, the successive ratios $\sigma_k(1,m_h)$ satisfy $\displaystyle \lim_{m_h\to \infty} \sigma_k(1,m_h)=1$.
In the continuum limit, we then fit our lattice data to
\bea
\sigma(1,m_h)\;= 1\;+\dfrac{s_1}{m_h} \;+\dfrac{s_2}{m_h^2}\;.
\eea
Clearly, the problem of larger errors for large quark
masses is circumvented by the above interpolation formula because $\sigma(1,\infty)=1$ ensures that the data with larger error bars become practically irrelevant in the fit.\\
Finally, after the interpolation to $\sigma(1,m_h)$, we obtain $\gsone$ at the physical $b$ quark mass by using a product of $\sigma_k(1)$ factors:
%\[
$
\gsone\,\equiv\,\mathscr{G}_s(1,m_h=m_b) \, \equiv \underbrace{\mathscr G_s(1,m_h=m_c)}_{=1} \,\sigma_0 \,\sigma_1 \cdots \sigma_7 \cdots \sigma_8\,,
$
%\]
leading to
\bea
\label{eq:gsone}
\gsone\,=\, 1.073(17) \;\; (\beta_k^s=0) \,,\qquad\qquad \gsone\,=\, 1.052(46) \;\; (\beta_k^s\neq 0) \,.
\eea
The result on the left, which is more accurate, agrees with the only existing unquenched Lattice QCD estimate, obtained for the light non-strange spectator quark~\cite{Okamoto:2004xg}.\\
Instead of starting from $\mathscr G_s(1,m_h=~m_c)$, we could have started from a $k < 9$, computed $\mathscr G_s(1,m_h=\lambda^{(k+1)}m_c)$ in the continuum limit, and then applied $\sigma_{k+1}\cdots \sigma_8$ to reach the physical $b$-quark mass. For example, by taking $k=3$, we obtain
\bea
\gsone \,=\, \mathscr G_s(1,\lambda^4 m_c) \, \sigma_4 \sigma_5 \cdots \sigma_8 \,=\,1.059(47) \qquad (\beta_k^s\neq 0)\, ,
\eea 
fully consistent with Eq.~\eqref{eq:gsone}.
%%%--------------------------------------
\section{The scalar and the tensor form factors}
%%%--------------------------------------
\noindent Latest experimental results by the BaBar Collaboration
for the ratio  $R(D)$ of the branching fractions $\mathscr{B}(B\to~D \mu\bar\nu_\mu)$
and $\mathscr{B}(B\to D \tau \bar\nu_\tau)$ suggest a disagreement with respect to the SM prediction~\cite{Aubert:2007dsa}. This discrepancy, which is around $2\sigma$,
might provide us with a first evidence for New Physics (NP) effects in semitauonic $B$ decay~\cite{BDpapers,BKT}.\\
In the models with two Higgs doublets (2HDM), the charged Higgs boson can mediate the tree level processes, including $B\to D\ell \bar\nu_\ell$,
and considerably enhance the coefficient multiplying the scalar form factor in the decay amplitude. By estimating the scalar form factor, involved in the SM theoretical prediction of these branching fractions, we can
interpret the discrepancy between the experimentally measured $R(D)$ and its theoretical prediction within the SM. Moreover, in the models of physics Beyond Standard Model (BSM) in which the tensor coupling to a vector boson is allowed, a third form factor might become important. 
In the present study, we make the first Lattice QCD estimate of the tensor form factor $F_T$  which, in the $B_s$ rest frame, is defined via
\bean
\langle D(\vec k)\vert T_{0i} \vert B (\vec 0)\rangle =  {{-2i\, m_{B_s}\, k_i} \,  \over m_{B_s}+m_{D_s}}\;F_T(q^2)\,, \quad \text{where} \quad T_{\mu\nu}=\bar c \sigma_{\mu\nu}b\,.
\eean
We focus on the determination of the ratios
\bean
 R_0(q^2)\,= \dfrac{F_0(q^2)}{F_+(q^2)}  \qquad {\text{and}} \qquad  R_T(q^2)\,= \dfrac{F_T(q^2)}{F_+(q^2)}\,,
\eean
at different values of $q^2 \lesssim q^2_{\rm max}$, but near the zero recoil (cf. Eq.~\eqref{nonzero}).\\

%which enter the differential decay rates of $B_s\to D_s$ process. Let us discuss the two cases.\\
%at different non zero recoils
\noindent\underline{$R_0(q^2)$:} we consider in this case ratios of $R_0$ computed at successive heavy quark masses and with a given value of $w$, namely
\bean
\displaystyle \Sigma^0_{(k)} (w,m_h^{(k)},a^2) \;= \dfrac{R_0(w,\, m_h^{(k+1)}, \,a^2 )}{ R_0(w,\, m_h^{(k)}, \, a^2 )}\,.
\eean
In order to obtain the continuum values $\sigma^0(w,m_h)$ of $\displaystyle \Sigma^0_{(k)} (w,m_h^{(k)},a^2)$, we apply
the continuum extrapolation by using an expression analogous to the one given in Eq.~\eqref{cont.extrap.}. As in the previous section we observe that our data do not show a significant dependence on the sea quark mass nor on the lattice spacing.\\
Using the HQET mass formula for $m_{B_s,D_s}$~\cite{neubert}, and
knowing that $h_+(w)$ scales as a constant with the inverse heavy quark mass, we deduce that the heavy quark interpolation can be performed by using~\cite{Atoui:2013zza}
\bea\label{fit:0plus}
\sigma^0(w,m_h) =  \frac{1}{\lambda} +  \frac{{s^{\prime}}_1(w)}{m_h} +  \frac{{s^{\prime}}_2(w)}{m_h^2} \,.
\eea
The physical value of the ratio $R_0(w)$ is then obtained by
%\[
$R_0(w,m_b) \,=\, R_0(w, \lambda^{k+1}m_c) \, \sigma_k^0(w) \cdots \sigma_8^0(w)$ ,{where} $ \sigma^0_k(w) \equiv \sigma^0(w,m_h^{(k)})$.
%\]
%%----------------------------------------
\begin{figure}[here!]
\subfigure[]
{
{\includegraphics[scale=0.4,angle=-0]{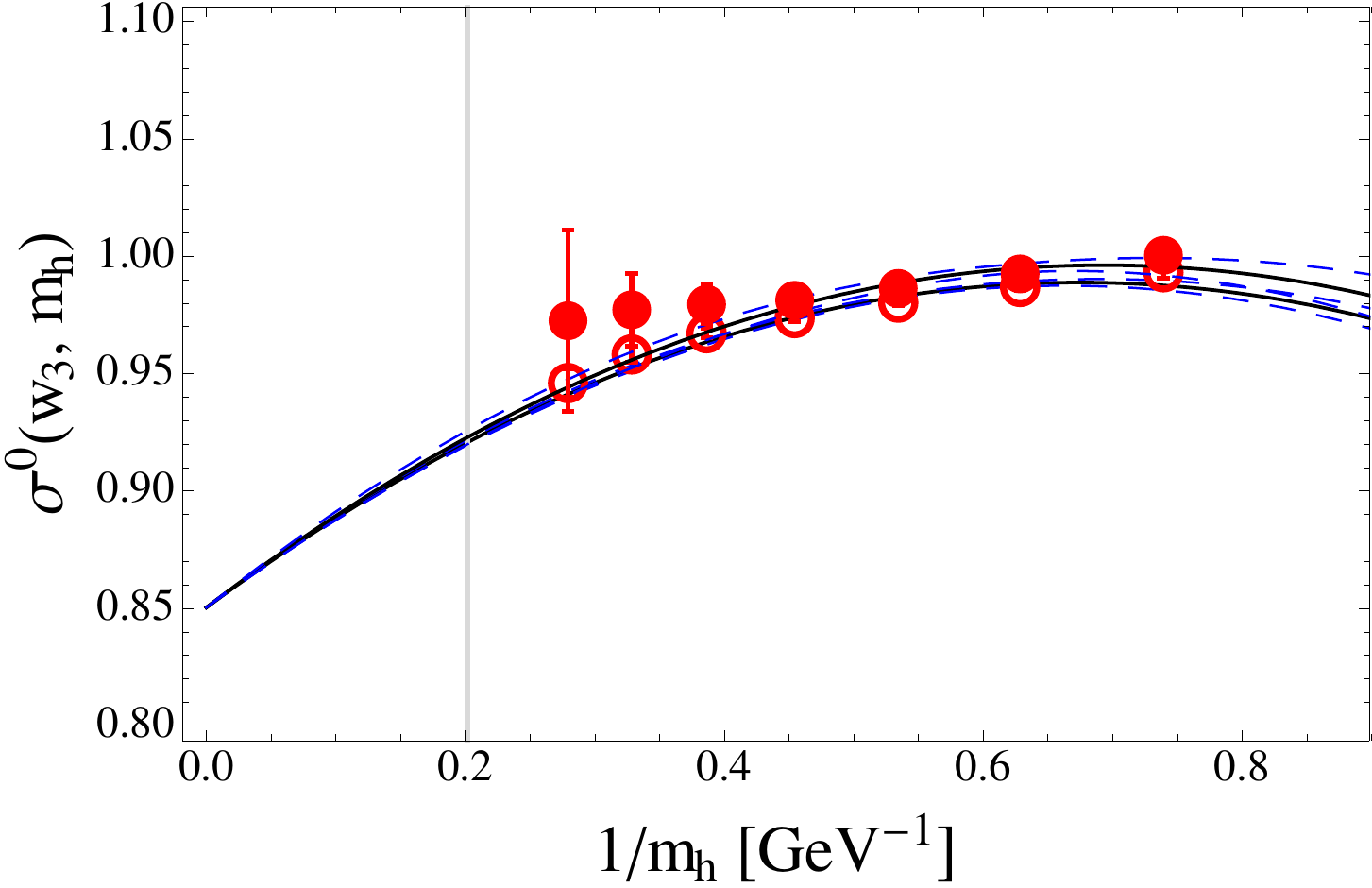}}
\label{fig:Sigma0plus}
}
\hspace*{1.5cm}
\subfigure[]
{
{\includegraphics[scale=0.4,angle=-0]{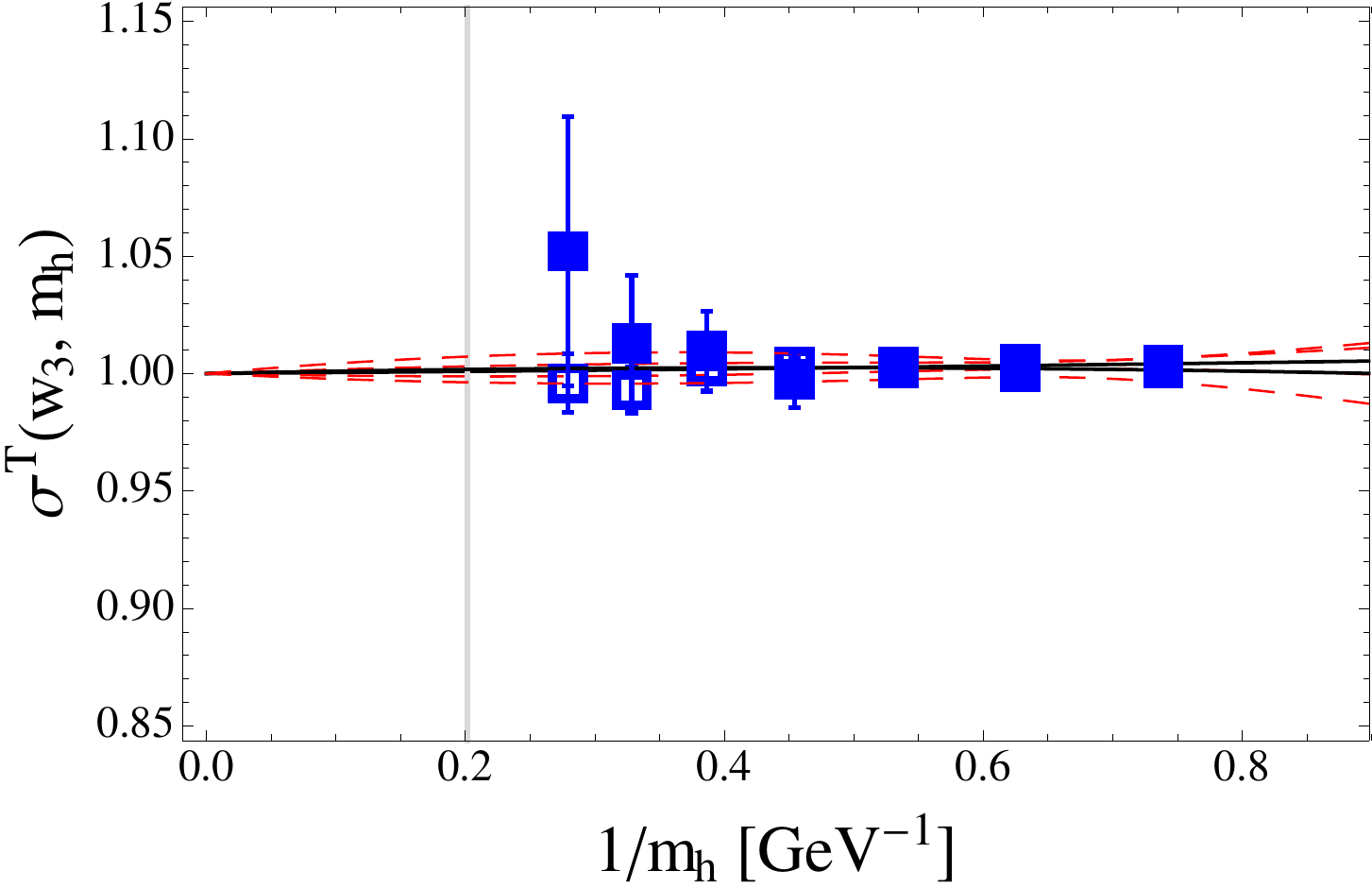}}
\label{fig:sigmat}
}
\centering
\caption{\label{fig:02} {\sl{(a) $\sigma^0(w)$ as a function of the inverse of the heavy quark mass, for $w=w_3=1.016$; (b) Same as (a) but for $\sigma^T(w)$. We see that the data obtained by assuming the independence of $R_0$ and $R_T$ on the sea quark mass (empty symbols) scale better than the results obtained by allowing a linear dependence on the sea quark mass (filled symbols).}}}
\end{figure}
%%-------------------------------------
\noindent In Fig.~\ref{fig:Sigma0plus}, we illustrate the ratio $\sigma^0$ for one specific value
of $w$ showing the data obtained by assuming both the dependence and the independence of $R_0(w)$ on the sea quark mass.\\% In Table~\ref{tab:03}, we present the physical values of $F_0(q^2)/F_+(q^2)$ as obtained starting from $k=3$.\\

\noindent\underline{$R_T(q^2)$:} 
%for the tensor operator $T_{\mu\nu}=\bar c \sigma_{\mu\nu} b$, the component which contributes to the tensor form factor, $F_T$, in the $B_s$ rest frame is $T_{0i}$
%\bean
%\langle D(\vec k)\vert T_{0i} \vert B (\vec 0)\rangle =  {{-2i\, m_{B_s}\, k_i} \,  \over m_{B_s}+m_{D_s}}\;F_T(q^2)
%\eean
It can be shown that the heavy quark behavior of the form factor $F_T(q^2)$ is similar to that of $F_+(q^2)$~\cite{Atoui:2013zza}. We again define the ratios computed at two successive quark masses that differ by a factor of $\lambda$,
\[
\displaystyle \Sigma^T_{(k)} (w,m_h^{(k)},a^2) \;= \dfrac{R_T(w,\, m_h^{(k+1)}, \,a^2 )}{ R_T(w,\, m_h^{(k)}, \, a^2 )}\,.
\]
As in the previous cases, we extrapolate $\Sigma^T_{(k)} (w)$ to the continuum limit and
observe that the result $\sigma_k^T(w,m_h)$ does not depend on the sea quark mass nor
on the lattice spacing (within
our error bars). The ratios $\sigma_k^T(w,m_h)$ are then fitted in the inverse heavy quark mass
%\bea\label{fit:tenplus}
$
\sigma^T(w,m_h) = 1 +  \frac{{s^{\prime\prime}}_1(w)}{m_h} +  \frac{{s^{\prime\prime}}_2(w)}{m_h^2}\, ,
%\eea
$
which is illustrated in Fig.~\ref{fig:sigmat} for $w=1.016$. The ratio $R_T(w)$ at $m_h=m_b$ is then obtained by using the following chain
$
R_T(w,m_b) \,=\, R_T(w, \lambda^{k+1}m_c) \, \sigma_k^T(w) \cdots \sigma_8^T(w)\, , \quad \text{where} \quad \sigma^T_k(w) \equiv \sigma^T(w,m_h^{(k)})\,.
$
We checked that our results for $R_T(w)$ obtained by starting from either $k=2$, $3$, or $4$, are completely consistent.
The results are given in Ref.~\cite{Atoui:2013zza}.
%in Table~\ref{tab:03} come from the choice $k = 3$.

%%----------------------------
%\begin{table*}[h!!]
%\centering
%{\scalebox{0.8}{
%\begin{tabular}{||c||c||c||} 
%\hhline{|t:=:t:=:t:=:t|}
%{\phantom{\huge{l}}} \raisebox{-.2cm} {\phantom{\huge{j}}}
%$w \; \l(q^2_{B_s\to D_s} {\text{GeV}}^2 \r)$  & {${F_0(q^2)\over F_+(q^2)}$} &{${F_T(q^2)\over F_+(q^2)}$}   \\ 
%\hhline{|:=::=::=:|}
%{\phantom{\huge{l}}} \raisebox{-.2cm} {\phantom{\huge{j}}}
%1.004 (11.46)  & 0.766(19)  & 1.076(68)   \\  
%{\phantom{\huge{l}}} \raisebox{-.2cm} {\phantom{\huge{j}}}
%1.016 (11.20)   & 0.781(24) & 1.062(76)    \\  
%{\phantom{\huge{l}}} \raisebox{-.2cm} {\phantom{\huge{j}}}
%1.036 (10.79)  &0.787(34)  & 0.975(94)   \\  
%{\phantom{\huge{l}}} \raisebox{-.2cm} {\phantom{\huge{j}}}
%1.062 (10.23)   & 0.825(59)  & 0.920(111)    \\  
%\hhline{|b:=:b:=:b:=b:|}
%\end{tabular}
%}}
%{\caption{  \label{tab:03} \sl
%Relevant physical results for $F_0(q^2)/F_+(q^2)$ and $F_T(q^2)/F_+(q^2)$ at various non zero recoil. Note that the renormalization scale for the tensor density is fixed to $\mu=m_b$ in the $\overline{\rm{MS}}$ scheme.}}
%\end{table*}
%%---------------------------
\noindent To our knowledge, the only existing result for $R_T(q^2)$ is the one of ref.~\cite{Melikhov:2000yu} for the non-strange case ($B \to~D~\ell\,\bar\nu_\ell$) in which the constituent quark model was used. Their result for the ratio $F_T/F_+$ was 1.03(1) and, in their work, this ratio was predicted to be a constant with respect to the momentum transfer $q^2$ (or the recoil $w$), but no reference to the renormalization scheme or scale could have been made.\\
%The tensor form factor might become important in the models of physics BSM in which the tensor coupling to a vector boson is allowed.

\section{Conclusion}
\noindent We computed the form factor $\mathscr G_s(1)$ by using the twisted mass QCD on the lattice.
That form factor is necessary for the theoretical description of the $B_s\to D_s\ell \bar\nu_\ell$ decay in the Standard Model and with the massless lepton in the final state 
 $\ell \in \{ e, \mu\}$. In doing so, we implemented the method proposed in~\cite{blossier} that allows to reach the physical value through the interpolation
of suitable ratios computed with successive heavy ``$b$'' quark masses for which a value for $m_h\to \infty$ is fixed by symmetry.
Our final result is $\mathscr G_s(1)~=~1.052(46)$.\\
Moreover, following the same methodology and restricting our attention to the small recoil region, we have determined the ratio of the scalar form factor to the vector one and for the first time in LQCD the tensor form factor with respect to the vector one. Of several $w$'s, we quote
\[
\l . \dfrac{F_0(q^2)}{F_+(q^2)} \r |_{q^2=11.5 \gev^2} = 0.77(2) \quad {\rm and} \quad  \l . \dfrac{F_T(q^2)}{F_+(q^2)} \r |_{q^2=11.5 \gev^2} = 1.08(7)\,.
\]
The above results are important for the discussion of this decay in various scenarios of physics BSM.\\
The same analysis has been done for the case of the non-strange decay mode and the results are fully consistent with those present for the $B_s\to D_s \ell \bar \nu_\ell$, but with larger statistical errors. More details can be found in~\cite{Atoui:2013zza}.

%%%%%%%%%%%%%%%

\end{document}